\documentclass[12pt,a4paper,english,nofootinbib]{revtex4}
\usepackage{lmodern}
\usepackage{lmodern}

\usepackage[T1]{fontenc}
\usepackage[latin9]{inputenc}
\usepackage{babel}
\usepackage{mathrsfs}
\usepackage{amsmath}
\usepackage{amssymb}
\usepackage{esint}
\usepackage[unicode=true,pdfusetitle,
 bookmarks=true,bookmarksnumbered=false,bookmarksopen=false,
 breaklinks=false,pdfborder={0 0 1},backref=false,colorlinks=false]
 {hyperref}

\makeatletter

\@ifundefined{textcolor}{}
{%
 \definecolor{BLACK}{gray}{0}
 \definecolor{WHITE}{gray}{1}
 \definecolor{RED}{rgb}{1,0,0}
 \definecolor{GREEN}{rgb}{0,1,0}
 \definecolor{BLUE}{rgb}{0,0,1}
 \definecolor{CYAN}{cmyk}{1,0,0,0}
 \definecolor{MAGENTA}{cmyk}{0,1,0,0}
 \definecolor{YELLOW}{cmyk}{0,0,1,0}
}

\usepackage{babel}
\usepackage{babel}
\usepackage{babel}
\usepackage{babel}
\usepackage{babel}
\usepackage{babel}
\usepackage{babel}
\usepackage{babel}
\usepackage{babel}
\usepackage{babel}
\usepackage{babel}
\usepackage{babel}
\usepackage{babel}
\usepackage{babel}
\usepackage{babel}
\usepackage{babel}

\usepackage{babel}
\usepackage{graphicx}
\def\b{\begin{equation}}
\def\e{\end{equation}}

\@ifundefined{textcolor}{}{%
 \definecolor{BLACK}{gray}{0}
 \definecolor{WHITE}{gray}{1}
 \definecolor{RED}{rgb}{1,0,0}
 \definecolor{GREEN}{rgb}{0,1,0}
 \definecolor{BLUE}{rgb}{0,0,1}
 \definecolor{CYAN}{cmyk}{1,0,0,0}
 \definecolor{MAGENTA}{cmyk}{0,1,0,0}
 \definecolor{YELLOW}{cmyk}{0,0,1,0}
 }

\usepackage{latexsym}\usepackage{bm}

\makeatother

\begin{document}
\title{{\normalsize{}{}{}{}{}Bowen-York Model Solution Redux }}
\author{{\normalsize{}{}{}{}{}{}{}{}{}{}{}{}Emel Altas}}
\email{emelaltas@kmu.edu.tr}

\affiliation{Department of Physics,\\
 Karamanoglu Mehmetbey University, 70100, Karaman, Turkey}
\author{{\normalsize{}{}{}{}{}{}{}{}{}{}{}{}Bayram Tekin}}
\email{btekin@metu.edu.tr}

\affiliation{Department of Physics,\\
 Middle East Technical University, 06800, Ankara, Turkey}
\date{{\normalsize{}{}{}{}{}{{}{}{}{}\today}}}

\maketitle

Initial value problem in General Relativity is often solved numerically; with only a few exceptions one of which is the "model" solution of Bowen and York where an analytical form of the solution is available.  The solution describes a dynamical, time-asymmetric, gravitating system with mass and linear momentum. Here we revisit this solution and correct an error which turns out to be important for identifying the energy-content of the solution. Depending on the linear momentum, the ratio of the non-stationary part of the initial energy to the total ADM energy takes values between $[0, 0.592)$. This non-stationary part  is expected to be turned into gravitational waves during the evolution of the system to possibly settle down to a black hole with mass and linear momentum. In the ultra-relativistic case (the high momentum limit), the maximum amount of gravitational wave energy is $59.2 \%$ of the total ADM energy. We also give a detailed account of  the general solution of the Hamiltonian constraint.

\section{{\normalsize{}{}{}{}{}{}{}{}{}{}{}{}Introduction}}

In this era of  frequent observations of gravitational waves from black hole collisions \cite{merger} or collisions of other compact objects, numerical and analytical study of Einstein's equations for the prediction of the wave profile and the resulting spacetime is extremely important to interpret the data. Of course almost  all of the relevant work is numerical and obviously any analytical solution would be extremely valuable. There is one such exact solution that we shall call the "Bowen-York model solution" given in \cite{BY} which we study here to understand the energy content of this solution as well as how the solution is obtained and how it can be generalized. As there is an error in the original work for the model solution, it has not been clear up to now whether or not the initial data has some non-stationary energy that will be converted to gravitational waves as  the system evolves. Here we correct this and give a systematic approach to the solution of the Hamiltonian constraint under the assumed conditions.
But first we briefly describe the  initial value problem.

Assuming that the spacetime  is topologically $\mathscr{M}=\mathbb{R}\times\Sigma$, with  $\Sigma$ being a spacelike hypersurface, Einstein's equations
\begin{equation}
R_{\mu\nu}-\frac{1}{2}Rg_{\mu\nu}+\Lambda g_{\mu\nu}=\kappa T_{\mu\nu}, \hskip 1 cm 
\end{equation}
can be turned into an initial value problem, a dynamical system with $\Sigma$ being the Cauchy surface. [We shall work in the $G =1=c $ units and  $\kappa= 8 \pi$.] To specify the initial data on the hypersurface, the spacetime metric can be decomposed as \cite{ADM}
\begin{equation}
ds^{2}=(N_{i}N^{i}-N^{2})dt^{2}+2N_{i}dtdx^{i}+\gamma_{ij}dx^{i}dx^{j},
\hskip 1 cm i, j \in (1,2,3),
\label{ADMdecompositionofmetric}
\end{equation}
with the lapse function $N=N(t,x^{i})$ and the shift vector $N^{i}=N^{i}(t,x^{i})$. Then one can take the Riemannian metric $\gamma_{ij}= \gamma_{i j}(t, x^j)$ which also lowers the spatial indices and the extrinsic curvature $K_{ij}= K_{ij}( t, x^k)$  to be the initial data on the Cauchy surface. The extrinsic curvature is defined as follows in a coordinate invariant manner: let $n$ be the unit normal to the spacelike hypersurface $\Sigma$, and $X,Y$ be two tangent vectors at that point to the hypersurface, and $\nabla$ be the spacetime-metric compatible connection, then $K(X,Y):=g(\nabla_{X}n,Y)$.  Of course by this definition, the extrinsic curvature is a purely spatial tensor and assuming  $D_{i}$ to be the covariant derivative compatible with $\gamma_{ij}$, one has explicitly 
\begin{equation}
K_{ij}=\frac{1}{2N}\Big(\dot{\gamma}_{ij}-D_{i}N_{j}-D_{j}N_{i}\Big), \hskip 1 cm \dot{\gamma}_{ij} = \frac{\partial}{\partial t}\gamma_{ij}.
\end{equation}
With these identifications, Einstein's equations yield, respectively, the Hamiltonian and momentum constraints on the hypersurface
$\Sigma$ as 
\begin{eqnarray}
 &  & -{}^{\Sigma}R-K^{2}+K_{ij}^{2}+2\Lambda-2\kappa T_{nn}=0,\nonumber \\
 &  & -2D_{k}K_{i}^{k}+2D_{i}K-2\kappa T_{ni}=0,~~~~~~~~~~~~~~\label{Einstein_c}
\end{eqnarray}
where $K:=\gamma^{ij}K_{ij}$ and $K_{ij}^{2}:=K^{ij}K_{ij}$; and the 
evolution equations for the spatial metric and the extrinsic
curvature as\footnote{We are writing the evolution equations just for completeness, we shall not use them in this work; their concise derivations can be found in the Appendix of \cite{our_dain_paper}. Moreover, in the same work one can also find how the linearized forms of the constraints  (\ref{Einstein_c}) also yield the evolution equations in the Fischer-Marsden form \cite{FM}. Hence the constraints play a double role.} 
\begin{equation}
\frac{\partial}{\partial t}\gamma_{ij}=2NK_{ij}+D_{i}N_{j}+D_{j}N_{i},
\end{equation}
\begin{equation}
\frac{\partial}{\partial t}K_{ij}=N\left(R_{ij}-{}^{\Sigma}R_{ij}-KK_{ij}+2K_{ik}K_{j}^{k}\right)+\mathscr{L}_{\overrightarrow{N}}K_{ij}+D_{i}D_{j}N,
\end{equation}
where $\text{\ensuremath{\mathscr{L}}}_{\overrightarrow{N}}$ is the
Lie derivative along the shift vector.  Note that ${}^{\Sigma}R_{ij}, {}^{\Sigma}R$ denote the intrinsic Ricci curvature and  the scalar curvature of the hypersurface, respectively. If we consider the vacuum case ($T_{\mu \nu} =0$) and with $\Lambda =0$, we have $R_{i j } = 0$. Having obtained a dynamical system for Einstein's equations, the way to proceed for finding solutions is clear, albeit analytically insurmountable without further assumptions. One should solve the Hamiltonian and momentum constraints to get viable initial data, then choose some lapse and shift functions to solve the evolution equations.  There are many approaches to these problems and the reader is invited to look at  the two excellent references \cite{Eric,Baumgarte}.
The method we shall consider is the one given by Bowen and York in their ground-breaking paper \cite{BY} where one can also find earlier relevant references of Misner \cite{Misner} and Brill-Lindquist \cite{BL} as the pioneers of exact solutions of the constraints for multiple black holes.

\section{BOWEN-YORK INITIAL DATA}

Following \cite{BY}, let us concentrate on the constraints (\ref{Einstein_c}) in a vacuum and with $\Lambda=0$.
Furthermore, assume that the Cauchy surface $\Sigma$ is conformally flat 
\begin{equation}
\gamma_{ij}=\psi^{4}f_{ij}, \hskip 1 cm \psi >0,
\end{equation}
with $f_{ij}$ denoting the flat metric in some coordinates. The inverse metric is
$\gamma^{ij}=\psi^{-4}f^{ij}$. The (physical) extrinsic curvature can be chosen in terms of a trial one as $K_{ij}=\psi^{-2}\hat{K}_{ij}$ such that one has 
$K_{i}^{j}=\psi^{-6}\hat{K}_{i}^{j}$ and $K^{ij}=\psi^{-10}\hat{K}^{ij}$.  Conformal flatness of the Cauchy surface simplifies the problem a lot, but it is not sufficient: one also assumes that it is a maximally embedded hypersurface in spacetime which boils down to setting the trace of the extrinsic curvature to zero
\begin{equation}
K=0.
\end{equation}
Denoting $\hat{D}_i$ to be the flat-metric compatible connection ({\it i.e.} $\hat{D}_i f_{ jk}=0$), then one obtains the intrinsic Ricci curvature of the hypersurface to be 
\begin{equation}
^{\varSigma}R_{ij}=-2\psi^{-1}\hat{D}_i\hat{D}_j\psi+6\psi^{-2}\hat{D}_i\psi\hat{D}_j\psi-2f_{ij}\psi^{-1}\hat{D}_k\hat{D}^k\psi-2f_{ij}\psi^{-2}\hat{D}_k\psi\hat{D}^k\psi,
\end{equation}
and the scalar curvature to be 
\begin{equation}
^{\varSigma}R=-8\psi^{-5}\hat{D}_i\hat{D}^i\psi.
\end{equation}
Then the Hamiltonian constraint, $^{\varSigma}R^{2}-K_{ij}^{2}=0$, 
becomes 
\begin{equation}
\psi^{7}\hat{D}_i\hat{D}^i\psi=-\frac{1}{8}\hat{K}_{ij}^{2},
\label{elliptic0}
\end{equation}
while the momentum constraint decouples from the conformal factor and simplifies a great deal:
\begin{equation}
\hat{D}^i \hat{K}_{ij} =0.
\label{trans}
\end{equation}
The solution strategy is then clear: one should solve the last equation and plug it to (\ref{elliptic0}) to solve for $\psi$. Out of all possible solutions to (\ref{trans}), Bowen-York chose the following 7-parameter ($p_i, a, {\cal{J}}_i$) solution on $ \mathbb{R}^3 -\{0\}$:
\begin{eqnarray}
 &  & \hat{K}_{ij}=\frac{3}{2r^{2}}\Big(p_{i}n_{j}+p_{j}n_{i}+(n_{i}n_{j}-f_{ij})p\cdot n\Big)+\epsilon\frac{3a^{2}}{2r^{4}}\Big(p_{i}n_{j}+p_{j}n_{i}+(f_{ij}-5n_{i}n_{j})p \cdot n\Big)\nonumber \\
 &  & ~~~~~~~~+\frac{3}{r^{3}}{\cal{J}}^{l}n^{k}\Big(\varepsilon_{kil}n_{j}+\varepsilon_{kjl}n_{i}\Big),\label{Bowen-York_extrinsiccurvature}
\end{eqnarray}
where $r\ne 0$ is the radial coordinate, $n^i$ is the unit normal on a sphere of radius $r$ (not to be confused with the unit normal to $\Sigma$); 
 $\epsilon=\pm 1$ and $p \cdot n=p^{k}n_{k}$. [As the equation is linear each bracketed part can be considered as a solution by itself; in fact see Beig \cite{Beig} for a more general solution.] The physical meaning of $p_i$ and ${\cal{J}}_i$ become clear if one assumes asymptotic flatness, {\it i.e.} $ \lim_{r\to\infty}\psi(r)=1+ {\mathcal{O}}( 1/r)$ so that the conserved total linear momentum of the Cauchy surface becomes
\begin{equation}
P_{i}=\frac{1}{ 8 \pi}\int_{S^2_\infty} dS\, n^j\, K_{ij}  = \frac{1}{ 8 \pi}\int_{S^2_\infty} dS \,n^j\, \hat{K}_{ij}  ,
\label{mom}
\end{equation}
while the total angular momentum reads
\begin{equation}
J_{i}=\frac{1}{ 16 \pi}\varepsilon_{i j k}\int_{S^2_\infty} dS\, n_l \,\Big ( x^j K^{k l} -  x^k K^{j l}\Big) = \frac{1}{ 16 \pi}\varepsilon_{i j k}\int_{S^2_\infty} dS\, n_l\, \Big ( x^j \hat{K}^{k l} -  x^k \hat{K}^{j l}\Big). 
\label{dad}
\end{equation}
Plugging (\ref{Bowen-York_extrinsiccurvature}) to (\ref{mom}) and (\ref{dad}), one arrives at
$P_i = p_i$ and $ J_i = {\cal{J}}_i$. So one has a gravitating, asymptotically flat system with a total linear and total angular  momentum given via these expressions. Observe that in these two conserved quantities the second term in (\ref{Bowen-York_extrinsiccurvature}) plays no role, namely the constant $a$ has not appeared yet, but that term will contribute to the ADM energy as we show below. To be able to compute the ADM energy, we have to be more specific about the asymptotic form of the scalar $\psi$. So let us assume (and this assumption must satisfy the constraint equations, and it does satisfy as we shall see below) 
\begin{equation} 
\lim_{r\to\infty}\psi(r)=1+ \frac{E}{2 r}+ {\mathcal{O}}( 1/r^2). \label{asymp1}
\end{equation}
Then the ADM energy 
\begin{equation}
E_{ADM} = \frac{1}{ 16 \pi}  \int_{S^2_\infty} dS\, n_i\, \Big ( \partial_j h^{ i j} - \partial_i h^j_j \Big) , 
\end{equation}
with $h_{i j} = (\psi^4 -1)\delta_{ ij}$ reduces to 
 \begin{equation}
E_{ADM} = -\frac{1}{ 2 \pi} \int_{S^2_\infty} dS\, n^i\, \partial_i \psi,   
\end{equation} 
and for (\ref{asymp1}) one has  $E_{ADM} = E$ as expected. But the all important question is to link $E$ to the other parameters ($p_i, a, {\cal{J}}_i$)  of the theory which we shall do below for the particular case of the Bowen-York model with zero angular momentum ${\cal{J}}_i$. For this case one has 
\begin{equation}
\hat{K}_{ij}^{2}=\frac{9}{2r^{4}}\left(\bigl(1+\frac{\epsilon a^{2}}{r^{2}}\bigr){}^{2}p^{2}+2\bigl(1-\frac{4\epsilon a^{2}}{r^{2}}+\frac{\epsilon^{2}a^{4}}{r^{4}}\bigr)(p\cdot n)^{2}\right),
\end{equation}
with $p^2 = p_i p^i$. So the Hamiltonian constraint  (\ref{elliptic0}) becomes 
\begin{equation}
\psi^{7}\hat{D}_i\hat{D}^i\psi=-\frac{9}{16r^{4}}\left(\bigl(1+\frac{\epsilon a^{2}}{r^{2}}\bigr){}^{2}p^{2}+2\bigl(1-\frac{4\epsilon a^{2}}{r^{2}}+\frac{\epsilon^{2}a^{4}}{r^{4}}\bigr)(p\cdot n)^{2}\right),
\label{elliptic}
\end{equation}
which is still a pretty complicated equation to solve. One can further simplify it by assuming that the linear momentum is in some direction, say the third direction $p_i = p \hat{z}$ and following \cite{BY} ignore the angular part (or set $\cos\theta =0$). Then the resulting equation reduces to a nonlinear  ODE:\footnote{Note that the corresponding equation (35) of the paper \cite{BY} (for $\epsilon =1$) is not correct and hence this leads to an incorrect interpretation of the resulting solution.}
\begin{equation}
\psi^{7}\frac{d}{dr}\left(r^{2}\frac{d}{dr}\psi\right)=-\frac{9p^{2}}{16r^{2}}\left(1-\epsilon\frac{a^{2}}{r^{2}}\right)^{2},\label{mod1}
\end{equation}
where $\psi=\psi(r) >0$. We would like to solve this equation for $r \in ( 0, \infty)$ with the following condition (for finite ADM energy as computed above)
\begin{equation} 
\lim_{r\to\infty}\psi(r)=1+ \frac{E}{2 r}+ {\mathcal{O}}( 1/r^2). \label{asymp2}
\end{equation}
Let us first observe that this asymptotic form is allowed by  (\ref{mod1}): as $r \rightarrow \infty$, one has $\psi^{7}\frac{d}{dr}\left(r^{2}\frac{d}{dr}\psi\right) \approx 0$ which is solved by  $\psi(r)=A+ \frac{B}{r}$.  We choose $A=1$ and $ B = E/2$ to obtain an asymptotically flat solution with a finite ADM energy $E >0$.

Before we embark on an attempt for the general solution, let us first study the particular solution of (\ref{mod1}) together with the boundary conditions (\ref{asymp2}) given by Bowen and York \cite{BY}; and correct some important numerical factors which are imperative in the interpretation of the solution. To guarantee the everywhere finiteness of the solution ( {\it i.e.}   the spatial metric and the extrinsic curvature), including $r =0$,
Bowen and York consider an "inversion-symmetric" solution: that is a solution which is intact (up to a possible sign change of the extrinsic curvature) under the Stokes-Kelvin transformations about the sphere at $r=a$. The inversion,  defined as $\bar r = a^2/r, \bar \theta = \theta, \bar \phi = \phi$ for $ r \ne 0$, acts as an isometry of the metric $\gamma_{i j}$. The effect of this isometry on the conformal factor can be found to be $ \psi (r,\theta, \phi) = \frac{ a}{r} \psi (\bar{r}, \bar{\theta} , \bar {\phi})$. Derivative of this relation yields a condition at the sphere:
\begin{equation}
\frac{ \partial \psi}{\partial r} + \frac{ 1}{ 2 a} \psi =0 \,\,\,  at \, \,\, r = a. \label{boundary1}
\end{equation} 
The solution (which is to be derived in the next section) satisfying (\ref{mod1}) and  (\ref{boundary1})  is 
\begin{equation}
\psi (r)=\left(1+\frac{2E}{r}+\frac{6a^{2}\epsilon}{r^{2}}+\frac{2a^{2}E\epsilon}{r^{3}}+\frac{a^{4}}{r^{4}}\right)^{1/4}, \label{modelsolution}
\end{equation}
if and only if the ADM energy is given as 
\begin{equation}
E=\sqrt{4a^{2}\epsilon+6 p^{2}},\hskip 1 cm \epsilon = \pm 1.
\label{dispersion1}
\end{equation}
Note that Bowen-York found the incorrect value (for the $\epsilon =1$ case) $E=\sqrt{4a^{2}+p^{2}}$. First let us observe that  in the case of $\epsilon =1$, for $ p=0$, we have the time-symmetric initial data with $K_{ij } =0$ and  the dispersion relation  becomes $E = 2 a$ with the solution (\ref{modelsolution}) reducing to  
\begin{equation}
\psi (r)= 1+ \frac{a}{r}. \label{Schwarz}
\end{equation}
This is the initial data for the Schwarzschild black hole together with the identification that the ADM mass of the black hole is $m = 2 a = E$. Clearly for the $p =0$ case $ \epsilon = -1$ does not make sense as it yields an imaginary ADM energy. So from now on, let us concentrate only in the  $ \epsilon = 1$ case. 

 In summary, the spatial metric of the Bowen-York model solution is 
\begin{equation}
ds^2_{\Sigma} = \left(1+\frac{2E}{r}+\frac{6a^{2}}{r^{2}}+\frac{2a^{2}E}{r^{3}}+\frac{a^{4}}{r^{4}}\right) ( dr^2 + r^2 ( d\theta^2 + \sin^2\theta d\phi^2 ) ),
\end{equation}
together with the extrinsic curvature (\ref{Bowen-York_extrinsiccurvature}) for ${\cal{J}}^{l}=0$. This solution has a total linear momentum $\vec{p} = p \hat{z}$ and ADM energy  (\ref{dispersion1}) with $\epsilon =1$.  This can be compared with the initial metric of the Schwarzschild black hole
\begin{equation}
ds^2_{\Sigma} = \left( 1+ \frac{a}{r}\right)^4 ( dr^2 + r^2 ( d\theta^2 + \sin^2 \theta d\phi^2 )),
\end{equation}
with a vanishing extrinsic curvature, momentum and energy $E = 2 a$.

The energy content of the solution is important to understand. Naively we can define the "non-stationary" energy  as the ADM energy of the dynamical solution minus the usual on-shell dispersion relation given as  $E_{0} =  \sqrt{m^2+p^2}$ with  $m = 2 a$ 
\begin{equation}
E_{non-stationary} := \sqrt{4a^{2}+6 p^{2}} - \sqrt{4a^{2}+p^{2}}.
\end{equation}
Defining the ratio of the non-stationary energy to the total energy as
\begin{equation}
\eta :=  1- \frac{E_{0}}{ E_{ADM} } =1-\frac{\sqrt{4 a^2+p^2}}{\sqrt{4 a^2+6 p^2}},
\label{eta}
\end{equation}
it takes values in the interval $\eta \in [0, \frac{\sqrt{6}-1}{\sqrt{6}})$ depending on the ratio of $p/m$ and for the ultra-relativistic case (  $p >> m$), $\eta$ approaches 0.592. Namely, in that limit about $59.2 \%$ of the initial energy is in the non-stationary form and one expects this energy to turn into gravitational radiation as time evolves. In the non-relativistic limit,  one has $E_{non-stationary} \approx \frac{5p^{2}}{2 m}$ and $\eta \approx \frac{5p^{2}}{2 m^2}$.

Note that if we had chosen $\cos\theta =1$,  (\ref{mod1}) would become 
\begin{equation}
\psi^{7}\frac{d}{dr}\left(r^{2}\frac{d}{dr}\psi\right)=-\frac{27p^{2}}{16r^{2}}\left(1-\epsilon\frac{a^{2}}{r^{2}}\right)^{2},\label{mod7}
\end{equation}
and the corresponding ADM energy relation becomes
\begin{equation}
E=\sqrt{4a^{2}\epsilon+\frac{9}{2} p^{2}},
\label{dispersion1}
\end{equation}
with $\eta$ as defined in (\ref{eta}) taking values as  $\eta \in [0, \frac{1}{3})$ and hence the maximum amount of gravitational radiation would be $33.3 \%$. 

Note that one could try to define a more refined version of the non-stationary energy of the initial data using the suggestions of Dain \cite{Dain} which were fully developed in \cite{Kroon, our_dain_paper} based on the notion of "approximate Killing symmetries" , that is approximate KIDS (Killing Initial data). But that computation would require the knowledge of not only the initial data  but also the lapse and the shift functions.   For the Bowen-York solution the shift function can be taken to be zero, but the lapse function is not unity, it must be found from the full Einstein's equations which is a non-trivial task which we shall come back to in another work.

\section{General solution of the Hamiltonian constraint}

Let us now try to give the general solution of (\ref{mod1})  together with the asymptotic flatness condition. For this purpose, we can first write it as a first order equation as follows. Let us first define $r:=a/u$ (we keep $\epsilon$ for now)
which then yields 
\begin{equation}
\psi(u)^{7}\frac{d^{2}}{du^{2}}\psi(u)=-\frac{9 p^{2}}{16a^{2}}\left(1-\epsilon u^{2}\right)^{2}.\label{mod2}
\end{equation}
So $u$ is dimensionless and takes values in the interval $u\in(0,\infty)$, but for the more relevant case of $\epsilon =1$, the inhomogeneous part vanishes at
$u =1$ and so one has to be careful with this point and divide the interval into two parts as $(0,1)$ and $(1,\infty)$.
The  asymptotic condition becomes 
\begin{equation}
\lim_{u\to0}\psi(u)=1+ \frac{E}{2 a}u+ {\mathcal{O}}( u^2),
\end{equation}
So we can recast  (\ref{mod2}) as\footnote{ For the case of $\epsilon =1$, assume that we are working in the interval $ u \in  (0,1)$. For the $(1,\infty)$ part of the interval, the form of the resulting equation will not change, but there will be some sign changes in the intermediate steps.} 
\begin{equation}
\frac{d^{2}}{du^{2}}\psi(u)=-c^{2}\psi(u)^{-3} {\cal {F}}\left(\frac{\psi(u)}{\sqrt{1-\epsilon u^{2}}}\right),\label{mod3}
\end{equation}
with $ {\cal {F}}(\chi):=\chi^{-4}$ and $c^{2}:=\frac{9 p^{2}}{16a^{2}}$. Let us now define a new function $\phi(u)$ in the following way \cite{Polyanin}
\begin{equation}
\phi(u):=\frac{\psi(u)}{\sqrt{1-\epsilon u^{2}}},
\end{equation}
then (\ref{mod3}) becomes 
\begin{equation}
(1-\epsilon u^{2})^{2}\phi''(u)-2\epsilon u(1-\epsilon u^{2})\phi'(u)-\epsilon\phi(u)=-c^{2}\phi(u)^{-3}  {\cal {F}}(\phi(u)),
\end{equation}
which, upon multiplying with $\phi'(u)$, reduces to
\begin{equation}
\left((1-\epsilon u^{2})^{2}\phi'(u)^{2}\right)'-\epsilon(\phi^{2}(u))'=-2c^{2}\phi(u)'\phi(u)^{-3}  {\cal {F}}(\phi(u)).
\end{equation}
Then integrating over $u$ yields 
\begin{equation}
(1-\epsilon u^{2})^{2}\phi'(u)^{2}-\epsilon\phi^{2}(u)+c_{1}=-2c^{2}\int d\phi(u)\phi(u)^{-3}  {\cal {F}}(\phi(u)),
\end{equation}
where $c_{1}$ is an integration constant. Since we know the function $  {\cal {F}}$, we
can integrate the right-hand side to get the desired first order equation
\begin{equation}
(1-\epsilon u^{2})^{2}\phi'(u)^{2}-\epsilon \phi^{2}(u)+c_{1}=\frac{c^{2}}{3}\phi(u)^{-6},\label{mod4}
\end{equation}
which is valid for both signs of $\epsilon$ in the full domain of $u$.
One can proceed to solve this equation, but at this stage it is a good idea to determine the integration constant $c_1$ using the boundary condition at $u=0$. We have
\begin{equation}
\phi(u=0) = 1, \hskip 1 cm \left.{\frac{d\phi}{ d u}}\right\rvert_{u=0} = \frac{E}{ 2 a},
\end{equation}
which yield
\begin{equation}
c_1 = \epsilon+\frac{c^2} {3} - \frac{E^2}{4 a^2}  =  \epsilon+\frac{3 p^2} {16 a^2} - \frac{E^2}{4 a^2},
\label{c1_equation}
\end{equation}
where in the second equality we inserted the value of $c^2$. Observe that in the Bowen-York's restricted, inversion symmetric solution at $u =1$, one has $c_1= 0$ and (\ref{c1_equation}) reduces to (\ref{dispersion1}).

We would like to solve (\ref{mod4}), but from now on  the discussion bifurcates for the sign choices of $\epsilon$.  For concreteness, and for its physical relevance, let us consider $ \epsilon = 1$, then the equation to be solved is the following
\begin{equation}
(1- u^{2})^{2}\phi'(u)^{2}- \phi^{2}(u)+c_{1}=\frac{c^{2}}{3}\phi(u)^{-6},
\label{e1eq1}
\end{equation} 
with 
\begin{equation}
c_1 = 1+\frac{c^2} {3} - \frac{E^2}{4 a^2}.
\end{equation}

In the region $u \in (0, 1)$, let us define 
\begin{equation}
\zeta := \frac{1}{2} \log \frac{ 1+ u}{1-u}, \hskip 1 cm    \zeta \in ( 0, \infty).
\end{equation}
Then (\ref{e1eq1}) becomes 
\begin{equation}
\left(\frac{d\phi}{d\zeta}\right)^{2}-\phi^{2}+c_{1}=\frac{c^{2}}{3}\phi^{-6}.
\end{equation}
Defining $\varphi(\zeta):=\phi(\zeta)^{2}$, it yields
\begin{equation}
\left(\varphi\frac{d\varphi}{d\zeta}\right)^{2}=4\varphi^{4}-4c_{1}\varphi^{3}+\frac{4c^{2}}{3},
\end{equation} 
with $\varphi(0)=1$. We can now separate and integrate it as 
\begin{equation}
\int_{1}^{\varphi(\zeta)}\frac{\varphi d\varphi}{\sqrt{\varphi^{4}-c_{1}\varphi^{3}+\frac{c^{2}}{3}}}=2\zeta.
\label{elliptic1}
\end{equation}
One can do this integral and find $\varphi$ as a function of $\zeta$ and trace back the steps to arrive at the conformal factor $\psi$. One can carry out  similar steps for the interval $u\in (1, \infty)$ and match the solution at $u=1$. That would constitute the most general solution of the differential equation. But the final expression after integrating the left-hand side of (\ref{elliptic1}) is in terms of the elliptic integrals of the first and third kinds and the result is not particularly illuminating to depict here in  its most general form. Instead we shall consider $c_1=0$, then the integral in (\ref{elliptic1}) gives
\begin{equation}
\left.\log \Bigg (\frac{\sqrt{\varphi^{4}+\frac{c^{2}}{3}}+\varphi^{2}}{\sqrt{\varphi^{4}+\frac{c^{2}}{3}}-\varphi^{2} }\Bigg)\right\rvert_1^{\varphi(\zeta)} =8\zeta. 
\end{equation}
Solving for $\varphi$ and tracing back all the intermediate redefinitions we made along the way, we arrive at the Bowen-York solution (\ref{modelsolution}) which seemed very ad hoc in the previous section and in the original work \cite{BY}. Of course this solution satisfies the inversion symmetry assumption and the boundary condition (\ref{boundary1}) hence the solution in the full domain is determined. 

Let us note that there is another particular value of $c_1$ for which the result of the integral (\ref{elliptic1}) can be written in terms of elementary functions. That value is  $c_1 = \frac{4}{3} c^{1/2}$, but the resulting expression yields an implicit function of   $\varphi$ in terms of $\zeta$. Let us also note that  for the $ \epsilon$ = -1 case, the following definition
\begin{equation}
\xi :=\mbox{ArcTan}(u),\hskip1cm\xi\in (\frac{\pi}{2},0),
\end{equation}
reduces (\ref{mod4}) to 
\begin{equation}
\left(\frac{d\phi(\xi)}{d\xi}\right)^{2}+\phi^{2}(\xi)+c_{1}=\frac{c^{2}}{3}\phi(\xi)^{-6},\label{mod5}
\end{equation}
with 
\begin{equation}
c_1 = -1+\frac{c^2} {3} - \frac{E^2}{4 a^2},
\end{equation}
and one proceeds exactly as in the other case.

\section{{\normalsize{}{}{}{}{}Conclusions}}

We have revisited the model solution of Bowen and York  for an initial gravitating system with a finite mass and linear momentum which is expected to settle down to a single non-spinning black hole as time evolves; and after correcting an error in the equation coming from the Hamiltonian constraint, we identified the amount of non-stationary energy in the data that will turn into gravitational radiation. Maximum amount of non-stationary energy, in the ultra-relativistic case approaches to $0.592$ of the total ADM energy of the system.   
 We have also given a detailed account of the general solution of the Hamiltonian constraint for the model problem, and  the solution turns out to be given in terms of elliptic functions. The steps involved in the general solution also makes the Bowen-York solution much more transparent.

\section{{\normalsize{}{}{}{}Acknowledgments}}

We would like to thank Ayse Karasu and Fethi Ramazanoglu for useful discussions.


\begin{thebibliography}{10}

\bibitem{merger} B.~P.~Abbott \textit{et al.} {[}LIGO Scientific
and Virgo Collaborations{]}, Observation of Gravitational Waves from
a Binary Black Hole Merger, Phys.\ Rev.\ Lett.\ \textbf{116}, no.
6, 061102 (2016).


\bibitem{BY}
J.~M.~Bowen and J.~W.~York, Jr.,
Time asymmetric initial data for black holes and black hole collisions,
Phys. Rev. D \textbf{21} (1980), 2047-2056.

\bibitem{ADM} R.~Arnowitt, S.~Deser and C.~Misner, The Dynamics
of General Relativity, Phys. \ Rev.\ \textbf{116}, 1322 (1959);
\textbf{117}, 1595 (1960); in \textit{{Gravitation: An Introduction
to Current Research}}, ed L. Witten (Wiley, New York, 1962).


\bibitem{our_dain_paper}
E.~Altas and B.~Tekin,
Nonstationary energy in general relativity,
Phys. Rev. D \textbf{101} (2020) no.2, 024035.


\bibitem{FM} A.~E. Fischer and J.~E. Marsden, The Einstein
evolution equations as a first-order quasi-linear symmetric hyperbolic
system. I., \newblock Comm. Math. Phys. \textbf{28}, 1 (1972).


\bibitem{Eric}
E.~Gourgoulhon,
3+1 formalism and bases of numerical relativity,
[arXiv:gr-qc/0703035 [gr-qc]].

\bibitem{Baumgarte}
T. Baumgarte and S Shapiro, Numerical Relativity: Solving Einstein's Equations on the Computer. Cambridge: Cambridge University Press (2010).

\bibitem{Misner}
C.~W.~Misner,
Wormhole Initial Conditions,
Phys. Rev. \textbf{118} (1960), 1110-1111.

\bibitem{BL}
D.~R.~Brill and R.~W.~Lindquist,
Interaction energy in geometrostatics,
Phys. Rev. \textbf{131} (1963), 471-476.

\bibitem{Beig}
R.~Beig,
Generalized Bowen-York initial data,
Lect. Notes Phys. \textbf{537} (2000), 55-69.

\bibitem{Dain}S. Dain, A New Geometric Invariant on Initial Data
for the Einstein Equations, Phys. Rev. Lett. \textbf{93}, 23, 231101
(2004).

\bibitem{Polyanin}
A. D. Polyanin and V. F. Zaitsev,
Handbook of Exact Solutions for Ordinary Differential Equations, Chapman \& Hall/CRC (2002).


\bibitem{Kroon}J. A. V. Kroon and J. L. Williams, Dain's invariant
on non-time symmetric initial data sets, Class. Quantum Grav. \textbf{34},
12, 125013, (2017).

\end{thebibliography}
\end{document}